\documentclass{ws-rv961x669}
\usepackage[utf8]{inputenc}
\usepackage{ws-rv-van}     % numbered citation/references (default)
\usepackage{ws-rv-thm}     % comment this line when `amsthm / theorem / ntheorem` package is used
\usepackage{algorithmic}
\usepackage{subfigure}     % required only when side-by-side / subfigures are used
\usepackage{threeparttable}

\makeindex
\begin{document}

\chapter[Spectrometers]{Spectrometers and Polyphase Filterbanks \\ in Radio Astronomy}\label{spec_chap}s

\author[D. Price]{Danny C. Price}

\address{Department of Astronomy, Campbell Hall 339 \\
University of California Berkeley,\\
Berkeley, CA 94720-3411 \\
dancpr@berkeley.edu}

\begin{abstract}
This review gives an introduction to spectrometers and discusses their use within radio astronomy. While a variety of technologies are introduced, particular emphasis is given to digital systems. Three different types of digital spectrometers are discussed: autocorrelation spectrometers, Fourier transform spectrometers, and polyphase filterbank spectrometers.  Given their growing ubiquity and significant advantages, polyphase filterbanks are detailed at length. The relative advantages and disadvantages of different spectrometer technologies are compared and contrasted, and implementation considerations are presented.

\end{abstract}

%\markright{Customized Running Head for Odd Page} % default is Chapter Title.

\body

\section{Introduction}\label{sec:intro}

A \emph{spectrometer} is a device used to record and measure the spectral content of signals, such as radio waves received from astronomical sources. Specifically, a spectrometer measures the power spectral density (PSD, measured in units of $\rm{WHz}^{-1}$) of a signal. Analysis of spectral content can reveal details of radio sources, as well as properties of the intervening medium. For example, spectral line emission from simple molecules such as neutral hydrogen gives rise to narrowband radio signals (Fig.~1), while continuum emission from active galactic nuclei gives rise to wideband signals.

There are two main ways in which the PSD---commonly known as \emph{power spectrum}---of a signal may be computed. The power spectrum, $S_{xx}$, of a waveform and its autocorrelation function,  $r_{xx}$, are related by the Wiener-Khinchin theorem. This theorem states that the relationship between a stationary (mean and variance do not change over time), ergodic (well-behaved over time) signal $x(t)$, its PSD, and its autocorrelation is given by
\begin{equation}
S_{xx}(\nu)=\int_{-\infty}^{\infty}r_{xx}(\tau)e^{-2\pi i\nu\tau}d\tau.
\label{eq:psd}
\end{equation}
where $\nu$ represents frequency, and $\tau$ represents a time delay or `lag'. The autocorrelation function is
\begin{equation}
r_{xx}(\tau)=\left\langle x(t)x(t-\tau)\right\rangle,
\end{equation}
where angled brackets refer to averaging over time. 

\begin{figure}[t]
 \centering
 \includegraphics[width=\textwidth]{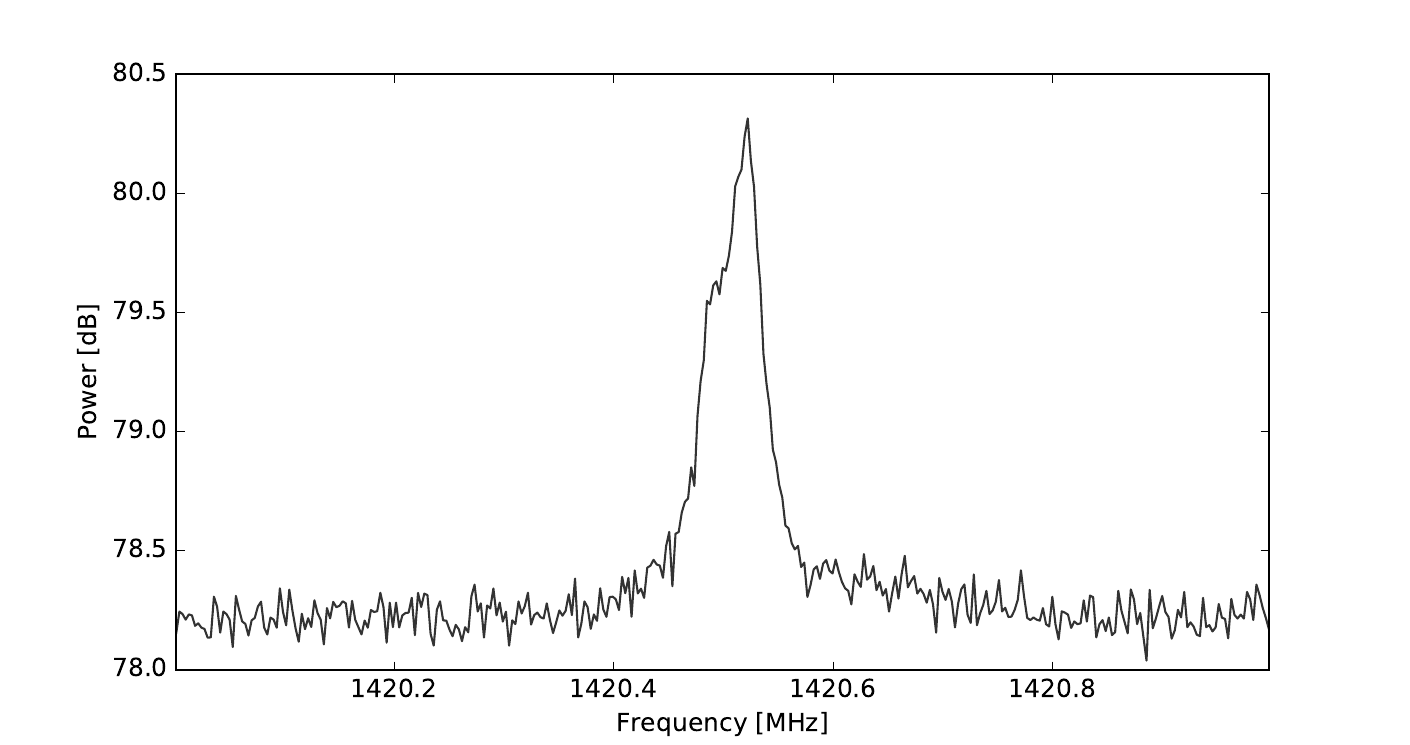}

 \label{fig:hydrogen}
 \caption{A galactic hydrogen 21-cm line emission profile, as measured using the Breakthrough Listen digital spectrometer system on the Robert C. Byrd Green Bank Telescope in West Virginia.}
\end{figure}

\begin{figure}[t]
 \centering
 \includegraphics[width=0.9\textwidth]{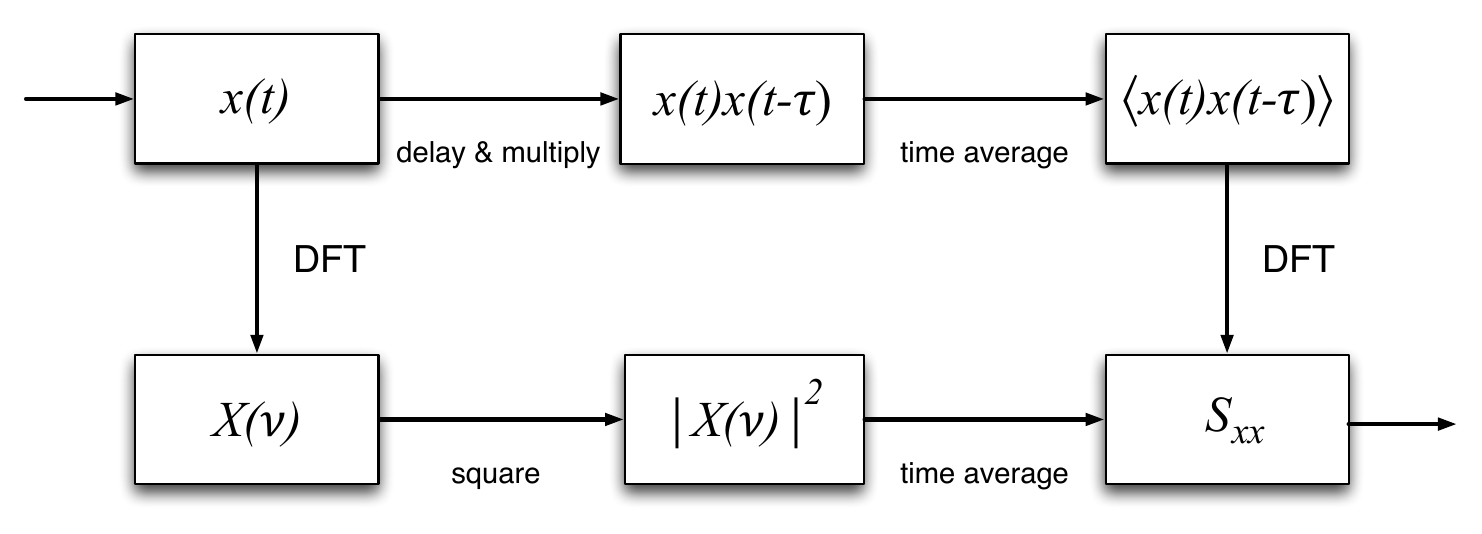}
 \caption{The two methods used to compute the PSD of a signal. The top path corresponds to an ACS system while the bottom corresponds to an FTF system. The two approaches are related by the Wiener-Khinchin theorem. \label{fig:wiener}  }
\end{figure}

\Eref{eq:psd} shows that that the autocorrelation function is related to the PSD by a Fourier transform. In the discrete case, the relationship becomes
\begin{equation}
S_{xx}(k)=\sum_{m=-\infty}^{\infty}\left\langle x(n)x(n-m)\right\rangle e^{-2\pi imk},\label{eq:discrete-wiener}
\end{equation}
which may be recognized as a discrete convolution. Here, the angled brackets average over time sample, $n$, and summation is performed over time lag, $m$. It follows from the convolution theorem that 
\begin{equation}
S_{xx}(k)=\left\langle \left|X(k)\right|^{2}\right\rangle ,\label{eq:discrete-pow}
\end{equation}
where $X(k)$ denotes the Discrete Fourier Transform (DFT) of $x(n)$:
\begin{equation}
X(k)=\sum_{n=0}^{N-1}x(n)e^{-2\pi ink/N}\label{eq:dft1}
\end{equation}
with $N \rightarrow \infty$. There are therefore two distinct classes of spectrometers: ones that approximate $S_{xx}(k)$ by firstly forming the autocorrelation, 
then taking a Fourier transform {\it à~la }
\eref{eq:discrete-wiener}, and those that first convert into the frequency domain to form $X(k)$ before evaluating \eref{eq:discrete-pow}. These two routes are shown diagrammatically in \fref{fig:wiener}. We will refer to these as autocorrelation spectrometers (ACS, \sref{sub:acs}), and Fourier transform filterbanks (FTF, \sref{sub:ftf}), respectively. Polyphase filterbank spectrometers (PFB, \sref{sub:pfb}) can be thought of as an  FTF with enhanced filter response. Note that because the DFT is an \emph{approximation} to the continuous Fourier Transform, ACS and FTF systems have different characteristics. 

% FIGURE
% FIGURE

\subsection{Analysis and synthesis filterbanks}

It is important to note the relationship between spectrometers, filters, and filterbanks. A \emph{filterbank} is simply an array of band-pass filters, designed to split an input signal into multiple components, or similarly, to combine multiple components. These are referred to as \emph{analysis} and \emph{synthesis} filterbanks, respectively. When applied to streaming data, a DFT can be considered an analysis filterbank, and an inverse DFT to be a synthesis filterbank. From this viewpoint, a spectrometer is simply an analysis filterbank, where the output of each filter is squared and averaged.

\subsection{Polarimetry}

Polarization is a key measurement within radio astronomy.\citet{BookTinbergenPolarim}  Although most astrophysical radio emission is inherently unpolarized, a number of radio sources---such as pulsars and masers---do emit polarized radiation, and effects such as Faraday rotation by galactic magnetic fields can yield polarized signals. A spectrometer that also measures polarization is known as a \emph{polarimeter} (or spectropolarimeter).

\subsubsection{Stokes parameters}

The Stokes parameters are a set of four quantities which fully describe the polarization state of an electromagnetic wave; this is what a polarimeter must measure. The four Stokes parameters, $I$, $Q$, $U$ and $V$, are related to the amplitudes of perpendicular components of the electric field:
\begin{eqnarray}
E_{x} & = & e_{x}(t)cos(\omega t+\delta_{x})\\
E_{y} & = & e_{y}(t)cos(\omega t+\delta_{y})
\end{eqnarray}
by time averages of the electric field parameters:
\begin{eqnarray}
I & = & \left\langle E_{x}E_{x}^{*}+E_{y}E_{y}^{*}\right\rangle \\
Q & = & \left\langle E_{x}E_{x}^{*}-E_{y}E_{y}^{*}\right\rangle \\
U & = & \left\langle E_{x}E_{y}^{*}+E_{y}E_{x}^{*}\right\rangle \\
V & = & i\left\langle E_{x}E_{y}^{*}-E_{y}E_{x}^{*}\right\rangle 
\end{eqnarray}
where $*$ represents conjugation. The parameter $I$ is a measure of the total power in the wave, $Q$ and $U$ represent the linearly polarized components, and $V$ represents the circularly polarized component. The Stokes parameters have the dimensions of flux density, and they combine additively for independent waves.

\subsubsection{Measuring polarization products}

In order to compute polarization products, a spectrometer must be presented with two voltage signals, $x(n)$ and $y(n)$, from a dual-polarization feed (i.e. a set of orthogonal antennas). With analogy to \eref{eq:discrete-pow}, we may form 
\begin{eqnarray}
S_{xx}(k) & =  \langle X(k)X^*(k)\rangle  & = \langle |X(k)|^2\rangle \label{eq:sxx1} \\
S_{yy}(k) & =  \langle Y(k)Y^*(k)\rangle  & = \langle |Y(k)|^2\rangle \\
S_{xy}(k) & =  \langle X(k)Y^*(k)\rangle  & \\
S_{yx}(k) & =  \langle Y(k)X^*(k)\rangle  & \label{eq:sxx4}
\end{eqnarray}
where in addition to measuring the PSD of $x(n)$ and $y(n)$, we also compute their cross correlations.  Note that while $S_{xx}$ and $S_{yy}$ are real valued, $S_{xy}$ and $S_{yx}$ are complex valued.

The four terms $\langle E_x E_x^*\rangle$, $\langle E_y E_y^* \rangle$, $\langle E_x E_y^* \rangle$, and $\langle E_y E_x^* \rangle$ are linearly related (by calibration factors) to the quantities in \eref{eq:sxx1} -- \eref{eq:sxx4} above. Combining these therefore allows for Stokes $I$, $Q$, $U$ and $V$ to be determined.

In order to focus on the fundamental characteristics of spectrometers, the remainder of this chapter details single-polarization systems that compute only $S_{xx}$. Nevertheless, the techniques and characterization approaches are broadly applicable to polarimetry systems.  See Chapter 9 of this volume for a more detailed treatment of radio polarimetry.

\subsection{Performance characteristics}

\begin{figure}
 \centering
 \includegraphics[width=\textwidth]{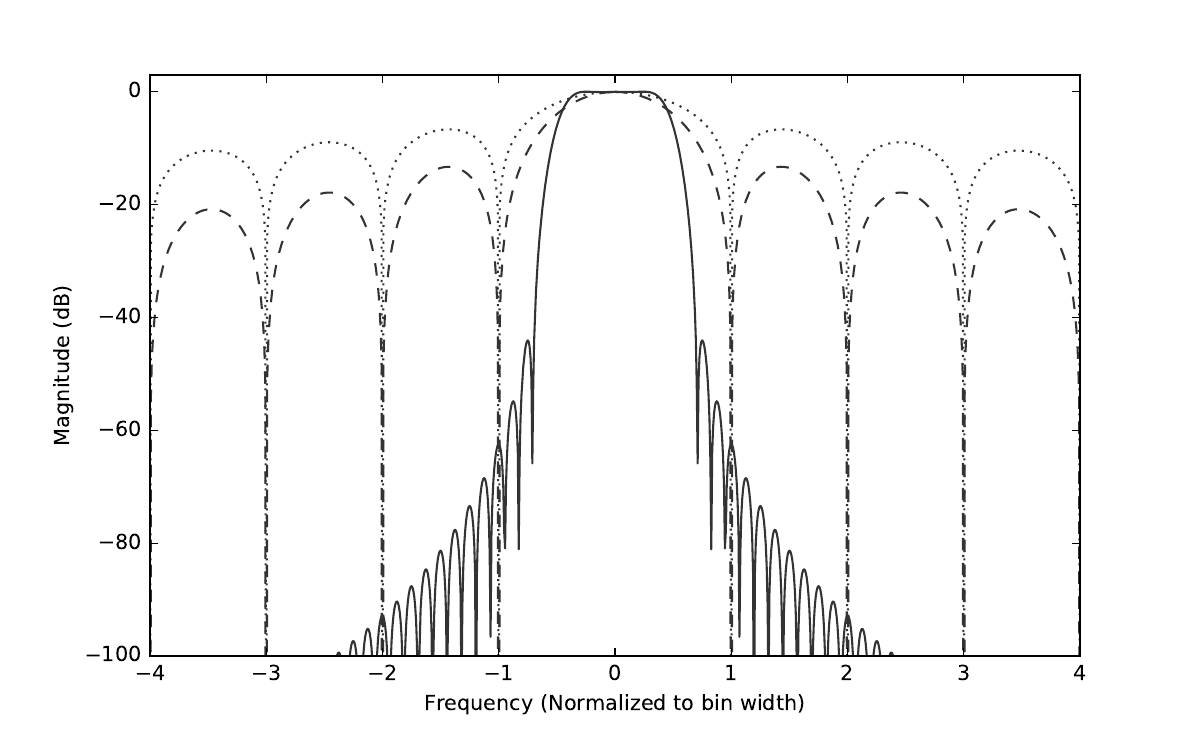}
 \label{fig:pfb_response}
 \caption{Comparison of the channel response of an ACS (dotted line), FTF (dashed line) and an 8-tap, Hann-windowed PFB (solid line).\label{fig:leakage}}
\end{figure}

Spectrometers operate over a finite bandwidth $B$, over which $N$ channels with bandwidth $\Delta\nu = B/N$ are computed.  With digital systems, channels may be evenly spaced with identical filter shapes. 

\subsubsection{Spectral leakage}

Ideally, each channel would have unitary response over $\nu_c \pm \frac{\Delta\nu}{2}$, where $\nu_c$ is the center frequency, with zero response outside this passband. In practice, this cannot be achieved; each channel has a non-zero response over all frequencies.  As such, a signal will `leak' between neighboring channels, known as spectral leakage.

\Fref{fig:leakage} compares the normalized filter response for ACS, FTF and PFB implementations. In the presence of strong narrowband signals, such as radio interference (RFI), spectral leakage is a major concern.

\subsubsection{Scalloping loss} 

\begin{figure}
 \centering
 \includegraphics[width=\textwidth]{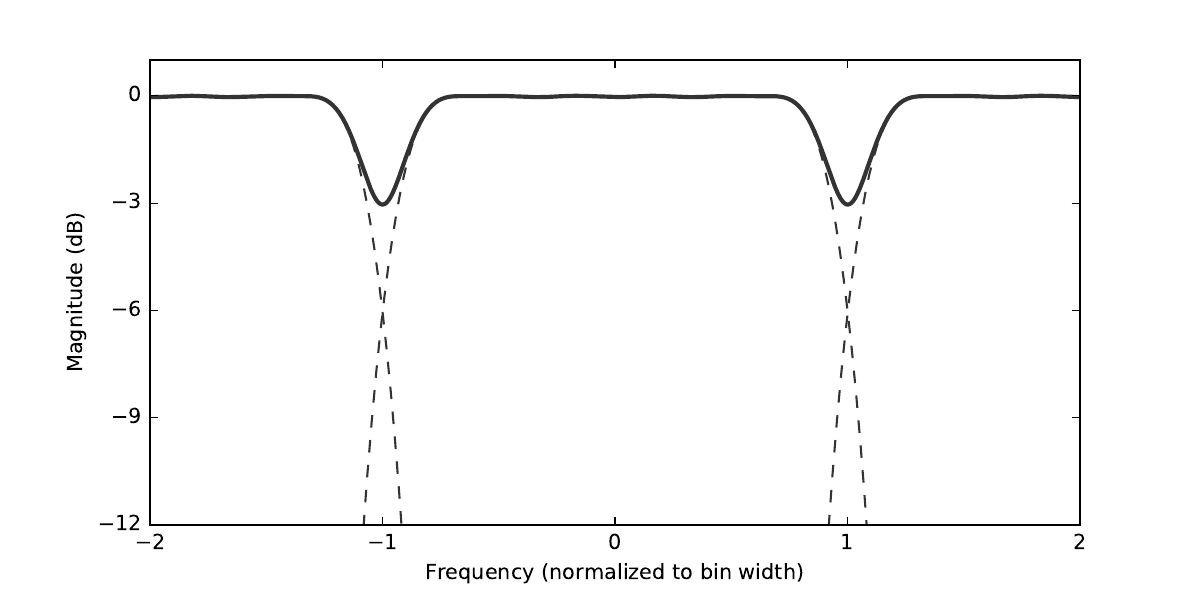}
 \caption{Example of scalloping loss between spectrometer channels. The dashed lines show the response of individual channels, while the black line shows the overall response.\label{fig:scalloping}}
\end{figure}

A related concern is that a channel's non-ideal shape will cause narrowband signals at channel edges to be attenuated, an effect known as scalloping loss (\fref{fig:scalloping}). Spectrometers are often designed such that neighboring channels overlap at their full-width at half-maximum points (FWHM), in which case the signal will be spread evenly over both channels. Wideband signals are not affected by scalloping. 

\subsubsection{Time resolution}\label{sub:time-res} 

Time resolution refers to the minimum period over which a spectrometer averages the data. For a spectrometer with $N$ channels over a bandwidth $B$, the time resolution is $t_{\rm{res}} = 2B/(RN)$, where $R$ is the length of the averaging window. Detection of transient phenomena, such as fast radio bursts and pulsars, require $t_{\rm{res}}$ to be as short as a microsecond, whereas integration lengths of several seconds, often averaged even further in post-processing, are common when observing faint sources.  

\subsubsection{Dynamic range}
Dynamic range refers to the span of input powers over which a spectrometer can operate nominally. The presence of RFI and the input bandwidth are the main drivers for dynamic range; see \sref{sub:dynamic-range}.

\section{Digital systems}

Digital signal processing (DSP) techniques are well-suited to applications such as filtering and forming filterbanks. As such, a majority of current-day spectrometers are based on digital technology. A basic understanding of DSP is required to fully understand digital spectrometers; there are several excellent introductory DSP texts available \cite{Lyons:2000uy, BookSmithFFT}. 

In the diagrams and equations in this chapter, the symbol $\otimes$ denotes multiplication of time samples; $\oplus$ denotes addition. The symbol $z^{-n}$ is used to denote a time delay of \emph{n} units, due to the relationship between time delay in a digital stream and the so-called $z$-transform.

\subsection{Digital sampling}

Digital sampling, or digitization, is the process of converting an analog signal to a digital one; devices known as analog to digital converters (ADCs) do this conversion. The two main characteristics of an ADC are its sample rate, $\nu_{\rm{s}}$, and the number of bits per sample, $n_{\rm{bits}}$.

\subsubsection{Nyquist sampling \label{sub:sampling}}
. 
The Nyquist Theorem---one of the most fundamental theorems within signal processing---states that a band-limited signal may be fully recovered when it is sampled at a rate that is twice the bandwidth, $\nu_{\rm{s}} = 2 B$. 
Sampling at the  Nyquist rate is referred to as \emph{critical sampling}, under the Nyquist rate as \emph{undersampling}, and sampling over the Nyquist rate as \emph{oversampling}. Sample rates may be increased by a process known as \emph{upsampling} and decreased by \emph{downsampling}, by using sample rate conversion filters. Here, we use the symbol $\downarrow D$ to denote downsampling by a factor \emph{D} and $\uparrow U$ for upsampling by a factor \emph{U}.

Undersampling a signal causes an effect known as \emph{aliasing} to occur, whereby different parts of a signal are indistinguishable from each other, resulting in information loss. Oversampling a signal does not increase the information content, but under certain circumstances is advantageous for reducing noise and/or distortion.

\subsubsection{Quadrature sampling}

\emph{Quadrature sampling}\cite{Lyons:2000uy} is the process of digitizing a band-limited signal and translating it to be centered about 0~Hz. A quadrature-sampled signal is complex valued, in contrast to real-valued Nyquist sampling. A quadrature-sampled signal has $\nu_{s} = B$; that is, each complex-valued sample is equivalent to two real-valued samples. Quadrature-sampled signals may have negative frequency components (i.e. below 0~Hz).

A Nyquist-sampled signal $x(n)$ centered at $\nu_0$ can be converted into a quadrature-sampled signal $x'(n)$ by multiplication with a complex phasor $e^{-2\pi i \nu_0 n}$:
\begin{equation}
	x'(n) = x(n) e^{-2 \pi i \nu_0 n},\label{eq:mix}
\end{equation}
which is known as quadrature mixing.

The $e^{-2\pi i \nu_0 n}$ term in \eref{eq:mix} is identical to that encountered in the DFT. Each channel of a DFT can be seen as quadrature mixing the input signal, applying a filter of width $B$, and then downsampling the signal to a rate $\nu_s=B$.

\subsubsection{Quantization efficiency \label{sub:quant}}

\begin{table}
	\tbl{Quantization efficiencies $\eta_Q$ for Nyquist sampling with different bit depths $N_{\rm{bits}}$. The value $\varepsilon$ is the threshold between quantized values, in units of the standard deviation of the signal.  Table modified from Ref.~\refcite{Thompson:2007p8886}.}
%	\begin{center}
	{\begin{tabular}{c c c c}
	\toprule
	$N_{\rm{bits}}$ & $N_{\rm{levels}}$ & $\varepsilon$ & $\eta_Q$ \\
	\colrule
	2 & 4   & 0.995  & 0.88115 \\
	3 & 8   & 0.586  & 0.96256 \\
	4 & 16  & 0.335  & 0.98846 \\
	5 & 32  & 0.188  & 0.99651 \\
	6 & 64  & 0.104  & 0.99896 \\
	7 & 128 & 0.0573 & 0.99970 \\
	8 & 256 & 0.0312 & 0.99991 \\
	\botrule
	\end{tabular}}
	\label{tab:quant_eff}
%	\end{center}
\end{table}

The earliest digital correlators \citep{Weinreb:1963p10042} used only two-level sampling (one bit), assigning a value of either +1 or -1. This scheme works remarkably well for weak, noise-dominated signals\footnote{That is, signals with probability distributions close to Gaussian.}: for Nyquist-sampled signals, a signal-to-noise ratio of $2/\pi=0.637$ that of the unquantized signal is achievable\citet{ThompsonMoranSwenson2004}. For 2-bit data, one can achieve 88\% quantization efficiency, which rises to 98\% for 4-bit data. Given these high values for low bit-widths, it is common to use bits sparingly within radio astronomy applications. A listing of quantization efficiencies\cite{Thompson:2007p8886} is given in \tref{tab:quant_eff}.

Achieving peak quantization efficiency relies on setting the threshold between quantized values optimally. In \tref{tab:quant_eff}, the threshold $\varepsilon$ is expressed in units of the signal's standard deviation $\sigma$. In order to leave headroom for interfering signals, one may deliberately set $\varepsilon$ larger than that optimal for signals with Gaussian probabilities to increase dynamic range.

\subsubsection{Dynamic range\label{sub:dynamic-range}}

For modern-day radio environments, RFI is the main driver of sampling bitwidth. RFI may be orders of magnitude stronger than an astronomical signal of interest, requiring a large dynamic range in the digitized waveform. If the maximum input power to an ADC is exceeded, an effect known as \emph{clipping} will occur, in which the waveform is distorted and spurious harmonics are introduced into the digitized waveform.

The theoretical maximum dynamic range of an ADC in decibels is given by 
\begin{equation}
	DR = 20\, \log_{10}(2^{n_{\rm{bits}}}) \approx 6.02 \, n_{\rm{bits}} \, .
\end{equation}
In practice, as ADCs are imperfect analog devices, their {\it effective number of bits} (ENOB) is lower than the number produced by the ADC. For example, an 8-bit ADC may have an ENOB of 7.5, resulting in a dynamic range of 45~dB.

\subsection{Windowing functions}

\begin{figure}[t]
 \centering
 \includegraphics[width=\textwidth]{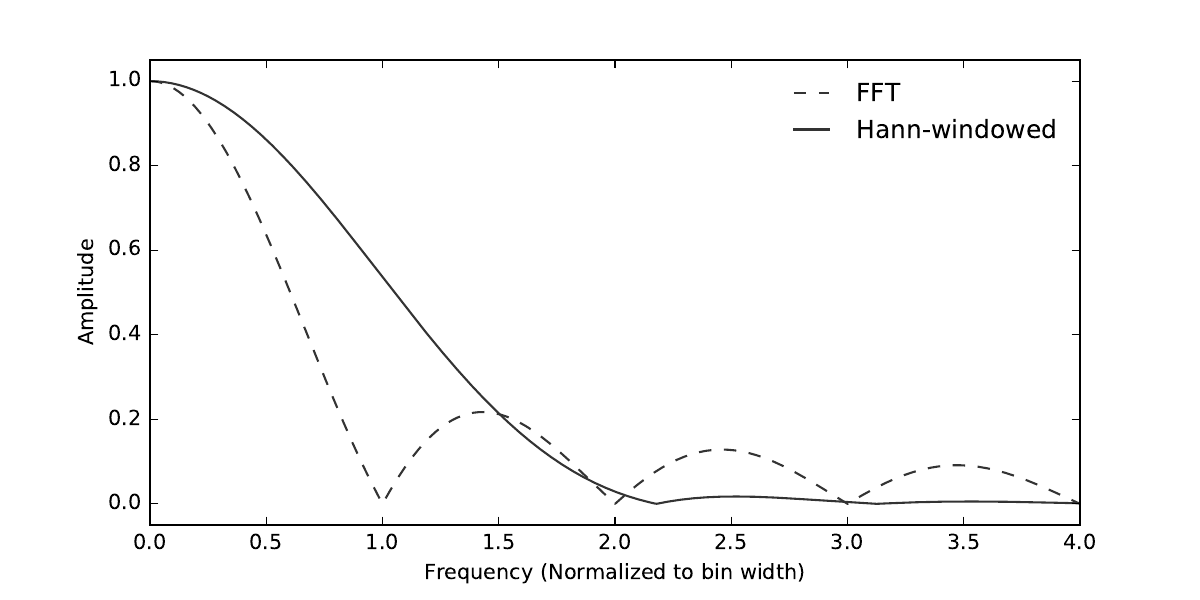}
 
 \caption{Amplitude response of the DFT (dashed line), compared to amplitude response of a Hann-windowed DFT (solid line). Applying a windowing function lowers sidelobes while broadening the channel response. \label{fig:fft_resp}}
\end{figure}

The DFT is computed over a finite number of samples, $N$, also known as the \emph{window length}. As the window length is not infinite, the response of the DFT is not perfect, resulting in spectral leakage (\fref{fig:leakage}). This can be understood if we consider that the DFT
\begin{eqnarray}
X'(k) & = & \sum_{n=0}^{N-1}x(n)e^{-2\pi ink/N} \\
     & = &  \sum_{n=-\infty}^{\infty} \Pi (n)x(n)e^{-2\pi ink/N}\\
     & = & \mathcal{F}\{\Pi_N(n)\}*X(k) \label{eq:dft_tophat} \, ,
\end{eqnarray}
where $\mathcal{F}$ denotes the Fourier transform, and $\Pi(n)$ the rectangle (or tophat) function:
\begin{equation}
\Pi_N(n) = \begin{cases}
	0 & \mbox{if } n < 0 \\
	1 & \mbox{if } 0 \leq n \leq N-1 \\
	0 & \mbox{if } n > N-1, \\
\end{cases}	
\end{equation}
which is Fourier paired with the sinc() function\footnote{${\rm sinc}(x) \equiv \sin(x)/x$.  This is the same relationship as that between light passing through a single slit aperture and its far-field diffraction pattern.}. In other words, we can consider the finite length of the DFT as effectively convolving the perfect Fourier transform response $X'(k)$ with a sinc function. The undesirable peaks of the sinc function are referred to as \emph{sidelobes}.

\begin{figure}[b]
 \centering
 \includegraphics[width=\textwidth]{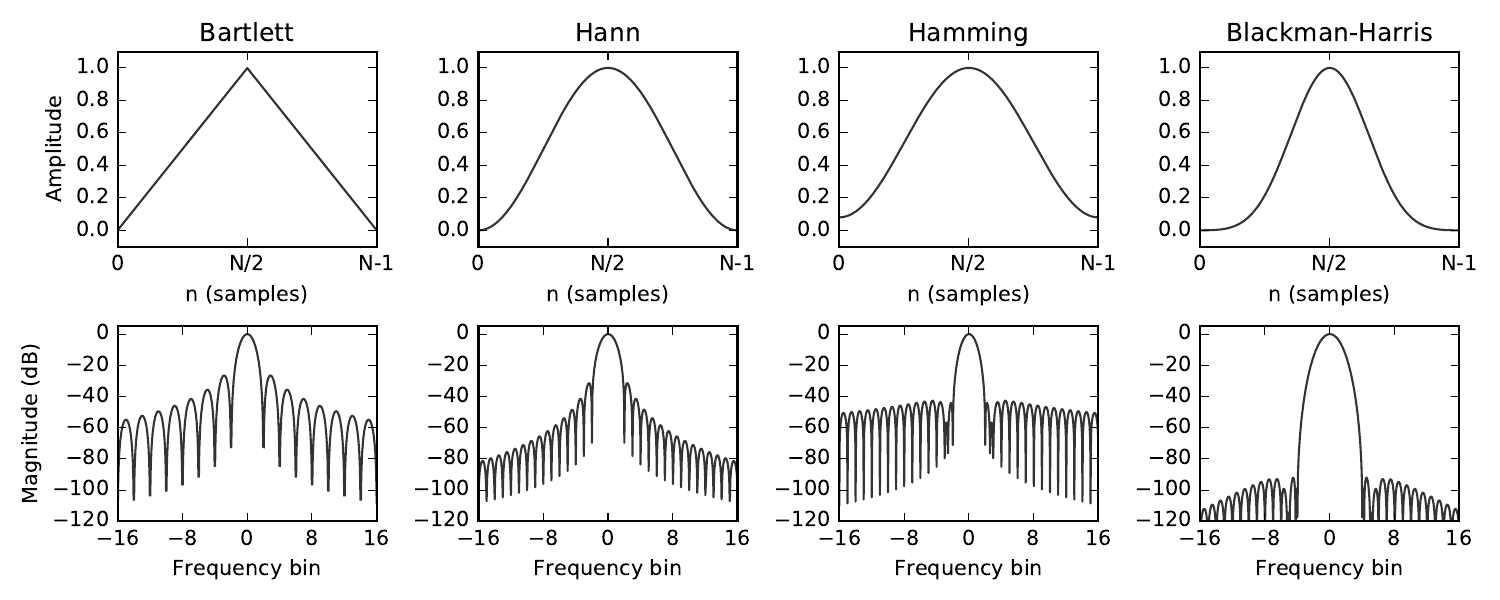}
 \caption{Four common windowing functions ($w(n)$, top) and their corresponding squared Fourier transforms ($|W(k)|^2$, bottom).  \label{fig:window_fns}}
\end{figure}

Windowing functions\citet{SvenGade1987} improve the response of a DFT, by somewhat mitigating sidelobe response at the expense of increasing the channel width. They are applied by multiplying the signal $x(n)$ by a weighting function, $w(n)$:
\begin{eqnarray}
X_w(k) & = & \sum_{n=0}^{N-1}w(n)x(n)e^{-2\pi ink/N} \\
       & = & W(k)*X(k).
\end{eqnarray}
The take-home message of all this is that DFT channels have a non-zero response outside their passband (\fref{fig:fft_resp}), and that applying a windowing function can improve their response. 

Windowing functions are also important in the design of digital filters (\sref{sec:filters}). Some common windowing functions and their frequency-domain magnitude responses are shown in \fref{fig:window_fns}; their functional forms are given in \tref{tab:window_fns}. The most appropriate windowing function is dependent upon application; for digital spectrometers, the Hamming and Hann windows are commonly applied.

\begin{table}[b]
	\tbl{Common windowing functions used in DFT filterbanks. Coefficients have been rounded to four significant digits. \label{tab:window_fns}}
	{\begin{tabular}{l l }
	\toprule
	Weighting function      & $w(n)$                        			\\
	\colrule
	Uniform  (rectangular)   &  1                           			\\
	Bartlett (triangular)    &  $1 - (|n| / (N-1)$               	\\
	\colrule
	\emph{General form:}    & $a_0 - a_1~cos(\frac{2\pi n}{N-1})$ \\
	Hann                     & $a_0=0.50 \quad a_1 = 0.50$ 	\\
	Hamming             & $ a_0 = 0.54 \quad a_1 = 0.46$   	\\
	\colrule
	
	 \emph{General form:} &$a_0 - a_1~cos(\frac{2\pi n}{N-1}) + a_2~cos(\frac{4\pi n}{N-1}) -
	 					   a_3~cos(\frac{6\pi n}{N-1}) $	 \\
	 Nutall          & $a_0=0.3558\quad a_1=0.4874\quad a_2=0.1442\quad a_3=0.0126$ \\
	 Blackman-Nutall & $a_0=0.3636\quad a_1=0.4892\quad a_2=0.1366\quad a_3=0.0106$ \\
	 Blackman-Harris & $a_0=0.3588\quad a_1=0.4883\quad a_2=0.1413\quad a_3=0.0117$ \\
	\botrule
	\end{tabular}}

\end{table}

\subsection{Finite impulse response filters}\label{sec:filters}

\begin{figure}
 \centering
 \includegraphics[width=0.8\textwidth]{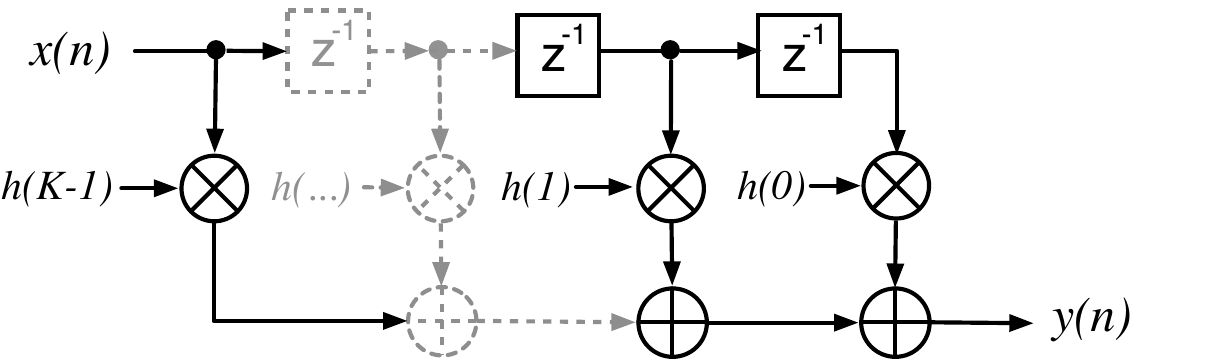}
 
 \caption{N-tap FIR filter block diagram. An FIR filter applies a weighted sum to the input sequence $x(n)$ to compute the filtered signal $y(n)$. \label{fig:fir}}
\end{figure}

A finite impulse response (FIR) filter is the windowed moving average of an input sequence $x(n)$. An FIR filter computes the sum 
\begin{equation} 
y(n)=\sum_{k=0}^{K-1}h(k)x(n-k),\label{eq:FIR-filter}
\end{equation}
where $y(n)$ is the output sequence, and $h(k)$ is a set of $K$ coefficients used for weighting. The upper summation bound, \emph{K}, is called the number of \emph{taps}. A streaming implementation of an FIR filter is shown in \fref{fig:fir}.

If downsampling a FIR filtered signal by $\downarrow D$, we only keep the outputs $n=rD$. 
In such cases it is more efficient to only compute the terms we wish to keep: 
\begin{equation}
y(rD)=\sum_{k=0}^{K-1}h(k)x(rD-k).\label{eq:FIR-filter-decimated}
\end{equation}
One way we can accomplish this is to use a polyphase decimating filter, which is discussed below.

\subsection{Polyphase FIR filters}\label{sub:pfir}

A common DSP technique is to decompose an input sequence $x(n)$ into a set
of $P$ sub-sequences, $x_{p}(n')$, each of which is given by 
\begin{equation}
x_{p}(n')=(\downarrow P)(z^{-p})x(n).
\end{equation}
This is known as polyphase decomposition\cite{Vaidyanathan:1990p6127}. As a simple example, even and odd decomposition of the signal $x(n)$ is achieved when $P=\mbox{2}$:
\begin{eqnarray}
x_{0}(n') & = & \left\{ x(0),x(2),x(4),...\right\} \\
x_{1}(n') & = & \left\{ x(1),x(3),x(5),...\right\} .
\end{eqnarray}
More generally, a signal may be decomposed into $P$ different `phases'.

Polyphase filter structures are often more efficient than standard finite impulse response filters when used in sample rate conversion. A $\downarrow P$ decimating FIR filter of length $K=MP$ can be constructed from $P$ discrete FIR filter \emph{branches}, each acting upon a different phase. The value $M$ is referred to as the number of \emph{polyphase taps} on each branch, such that
\begin{equation}
y(n')=\sum_{p=0}^{P-1}\sum_{m=0}^{M-1}h_{p}(m)x_{p}(n'-m),\label{eq:FIR-polyphase-filter}
\end{equation}
This is known as a decimating polyphase filter.

Decimating polyphase filter structures are far more efficient than standard FIR-based downsampling techniques. If $\downarrow D$ downsampling occurs after the moving average of \eref{eq:FIR-filter}, we are computing \emph{D} sums, but only keeping 1 in \emph{D} of these. This is inefficient; in contrast, \eref{eq:FIR-polyphase-filter} only computes values of interest. 

\subsection{The Fast Fourier Transform\label{sub:fft}}

The Fast Fourier Transform\cite{Cooley1965, BookBrighamFFT} (FFT) is a highly efficient algorithm for computing the DFT of a regularly-sampled signal. When applied over non-overlapping blocks of length $N$ of a time stream -- as done in FTF systems -- we may write the $r$-th output of a DFT as
\begin{equation}
X(k, rN)=\sum_{n=0}^{N-1}x(rN-n)e^{-2\pi ink/N}\label{eq:dft_ts}
\end{equation}
By comparison with \eref{eq:FIR-filter-decimated}, we recognize this as a bank of $N$ FIR filters, downsampled by $\downarrow N$. This is a key insight toward understanding DFT-based filterbanks: the DFT should be thought of as more than just a transformation from time to frequency domain.

To directly compute the DFT would require of order $O(N^2)$ operations, but the FFT algorithm reduces this to only $O(N \rm{log_2}$$N)$ operations. FFT implementations generally exhibit best performance when $N$ is a power of 2. 

For real-valued data, only $N/2$ channels are unique. FFT algorithms often exploit this for increased efficiency, by recasting the real-valued input data as complex values under-the-hood. The  $O(N \rm{log_2}$$N)$ performance of the FFT is a major driving factor for the adoption of FTF spectrometers over their ACS counterparts, which require $O(N^2)$ operations.

\section{Digital spectrometers}
As discussed in \sref{sec:intro}, there are two equivalent paths that may be used to compute the PSD of a signal, as shown in  \fref{fig:wiener}, referred to as ACS and FTF systems. As the DFT must be computed over a finite number of points, ACS and FTF systems have different characteristics. 

The first digital spectrometer used for radio astronomy was developed by Weinreb\citet{Weinreb:1963p10042} in 1963 -- two years before the FFT algorithm was introduced by Cooley \& Tukey\cite{Cooley1965}. This 1-bit ACS was used to observe the 18-cm wavelength hydroxyl (OH) absorption line in the spectrum of Cassiopeia A, providing the first evidence of OH in the interstellar medium \citep{Weinreb:1963p9992}. The first reference to FTF spectrometers for radio astronomy can be found in Chikada et. al.\citet{Chikada:1987p10044}; however FTF spectrometers did not enjoy widespread adoption until much later. The PFB architecture was first introduced by Schafer\cite{Schafer1973} in 1973 and expounded by Bellanger\cite{Bellanger:1976p7898} in 1976, but was not introduced for the purposes of radio astronomy spectrometry until 1991\cite{zimmerman1991, Duluk1992}. Bunton\cite{Bunton2000} further popularized the PFB within radio astronomy in 2000, suggesting its use in radio interferometer correlator systems. Given their growing ubiquity, PFB systems (which are essentially enhanced FTF spectrometers) are detailed at length in \sref{sub:pfb}.

Spectrometers do not compute the true PSD, $S_{xx}(k)$; rather, they compute an approximation, $S'_{xx}(k)$. Further, as a spectrometer has time resolution (\sref{sub:time-res}), the spectrometer output has a time dimension, i.e. $S'_{xx} = S'_{xx}(k, r)$, where $r$ is the integration number. 

\subsection{Autocorrelation spectrometers}\label{sub:acs}

In an ACS, the PSD is computed over a discrete range of $M$ delays:
\begin{eqnarray}
S'_{xx}(k) & = & \sum_{m=0}^{M-1}\left\langle x(n)x(n-m)\right\rangle e^{-2\pi imk} \\
          & = &  \sum_{m=-\infty}^{\infty}\Pi_N(n)\left\langle x(n)x(n-m)\right\rangle e^{-2\pi imk} \\
          & = & {\rm sinc}(k) * S_{xx}(k).\label{eq:acs_sinc}
\end{eqnarray}
That is, the finite summation causes convolution of the true PSD with a sinc() function. 

The spacing of lags (i.e. delays, $\tau$) in an ACS determines how much bandwidth it can process without aliasing occurring. The Nyquist criterion requires two taps per wave period at the highest frequency signal of interest, with the maximum lag $\tau_{\rm{max}}$ setting the spectral resolution, $\Delta\nu = \sim 1 / \tau_{\rm{max}}$.

\subsection{Fourier transform spectrometers}\label{sub:ftf}

FTF spectrometers compute the PSD of a signal by applying a DFT of length $N$ to an input signal, squaring the DFT output, then taking an average over time. From \eref{eq:discrete-pow} and \eref{eq:dft_tophat}, a FTF spectrometer computes

\begin{eqnarray}
	S'_{xx}(k) & = & \left\langle \left|X'(k)\right|^{2}\right\rangle \\
	    & = & \left\langle \left|{\rm sinc}(k)*X(k)\right|^{2}\right\rangle \\
	    & = & {\rm sinc}^2(k) * S_{xx}(k) \label{eq:ftf_sinc}
\end{eqnarray}
That is, the finite DFT summation bounds give rise to a convolution of the true PSD with a sinc$^2$() function.

For a signal with sampling rate $\nu_s=2B$, each DFT channel has a bandwidth $\Delta\nu \sim B/N$, and a quadrature-sampled output rate of $\nu_s/2N = B/N$. As mentioned in \sref{sub:sampling}, the DFT (\eref{eq:dft1}) may be thought of as the mixing of the input signal with a bank of oscillators, followed by an averaging with a square window function. 

\subsection{FTF and ACS comparison}\label{sub:ftf}

The FFT (\sref{sub:fft}) allows \eref{eq:dft1} to be evaluated in $O(N\log_2N)$ operations, or $\nu_s\log_2N$ when performed every $N$ samples. In comparison, an ACS requires $O(N^2)$ computations. For a spectrometer with a moderate $10^4$ channels, the FFT algorithm requires approximately $0.1\%$ as many operations as an ACS system.
 
ACS systems are more affected by spectral leakage than FTF systems (\fref{fig:leakage}), due to the sinc() convolution in \eref{eq:acs_sinc}, versus the sinc$^2$() convolution encountered in FTF systems (\eref{eq:ftf_sinc}). With current digital technology there is no compelling reason to implement an ACS spectrometer. Regardless, the earliest digital spectrometers were ACS based. Their prevalence in early systems can be explained by two reasons: for 1-bit data they can be implemented using simple boolean logic circuits; further, they pre-date the FFT algorithm.  
 
\subsection{Polyphase filterbanks}\label{sub:pfb}

A PFB is a computationally efficient implementation of a filterbank, constructed from an FFT preceded by a prototype polyphase FIR filter frontend.\citet{Schafer1973, Bellanger:1976p7898,  Harris2011} PFB-based spectrometers offer vastly lowered spectral leakage over both ACS and FTF architectures, with a modest increase in computational requirements.

The PFB exploits the fact that a lowpass filter with coefficients $h(k)$ can be converted into a quadrature bandpass filter with central frequency $\nu$ by multiplying the coefficients by $e^{i 2\pi \nu}$ . 

\begin{figure}
 \centering
 \includegraphics[width=\textwidth]{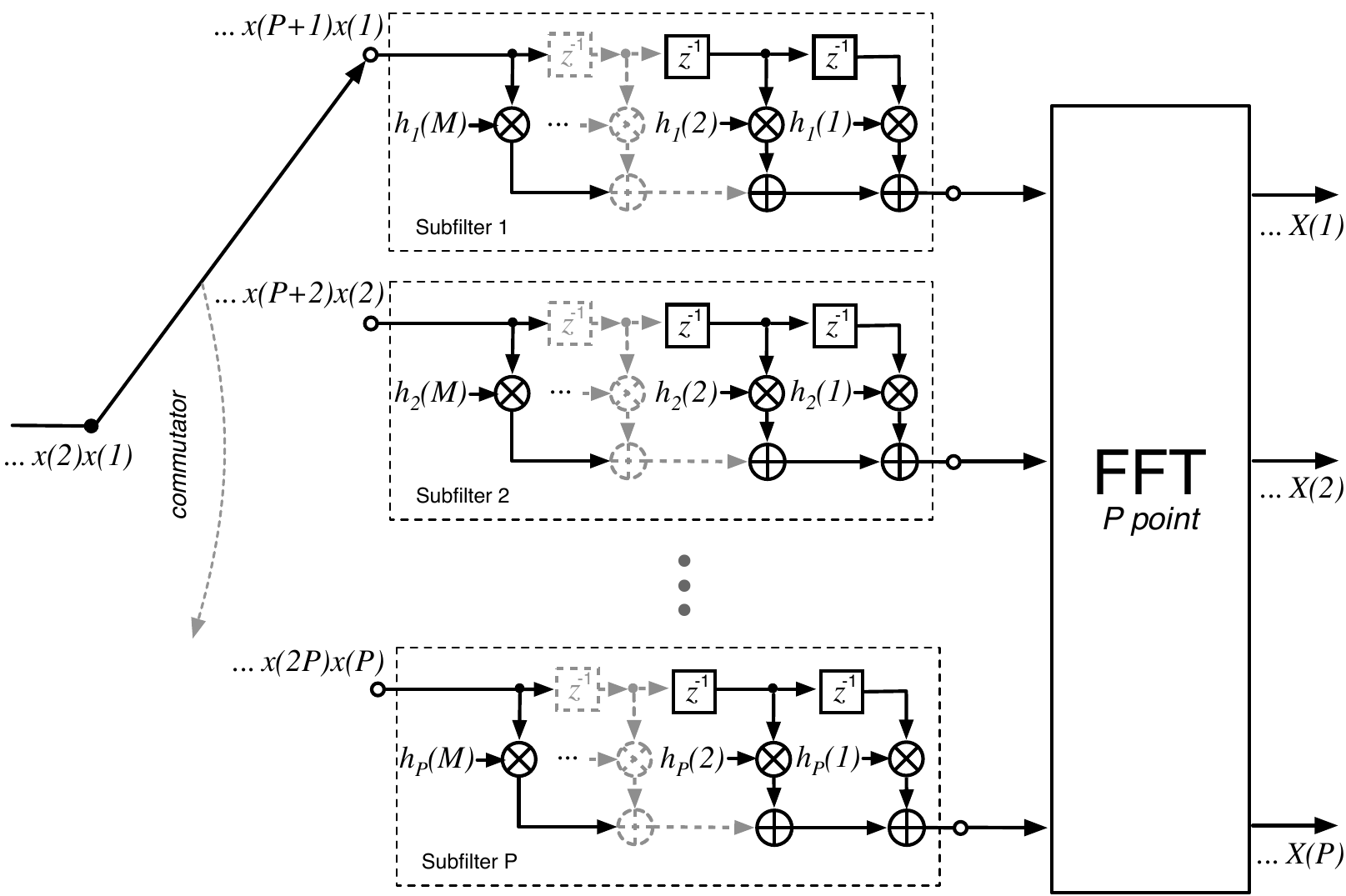}
 \caption{Polyphase filterbank streaming implementation. A PFB is formed when a polyphase FIR filter structure is combined with a DFT. This diagram is an alternative, but equivalent, representation to \fref{fig:pfb_chart}. Note that in this diagram, indices $m$ and $p$ run from 1 instead of 0. \label{fig:pfb_fir_fft}}
\end{figure}

Now, suppose we have implemented a decimating lowpass polyphase filter (\eref{eq:FIR-polyphase-filter}). The output of each branch is 
\begin{equation}
y_{p}(n')=\sum_{m=0}^{M-1}h_{p}(m)x_{p}(n'-m),
\end{equation}
where $h_{p}(m)$ are coefficients from our prototype lowpass filter. Normally, we would sum across the $P$ branches (i.e. over the sub-filters $y_p$) to construct $y(n')$, as in \eref{eq:FIR-polyphase-filter}.
Here is where we get tricky. If instead of just summing up the $P$ branches, we feed the branches (sub-filters) into a DFT with $P$ inputs, as in \fref{fig:pfb_fir_fft}, we then have 
\begin{eqnarray}
Y(k, n') & = & \sum_{p=0}^{P-1}y_{p}(n')e^{-2\pi ikp/P}\\
 & = & \sum_{p=0}^{P-1}\sum_{m=0}^{M-1}[h_{p}(m)e^{-2\pi ikp/P}]x_{p}(n'-m). \label{eq:pfb-final}
\end{eqnarray}
Comparing this form to \eref{eq:FIR-polyphase-filter}, we recognize that the output of this structure is equivalent to a set of $\downarrow P$ decimating polyphase filters, where the central frequency of each filter is shifted by an amount $p/P$; this is a polyphase filterbank. From here, the output is squared and time averaged to form the PSD. 

The order of summations in \eref{eq:pfb-final} is important; as written, it is more computationally efficient. The overhead of \eref{eq:pfb-final} over a windowed FTF\footnote{As an aside: a windowed FTF can be considered to be a one-tap PFB.} is an extra $(M-1)P$ operations, due to the polyphase FIR frontend. Generally, the number of taps $M\ll P$, so the increase in required operations is moderate. Extra memory is also required for buffering of the $M\times P$ time samples and filter coefficients.  .

\begin{figure}[t]
 \centering
 \includegraphics[width=1.0\textwidth]{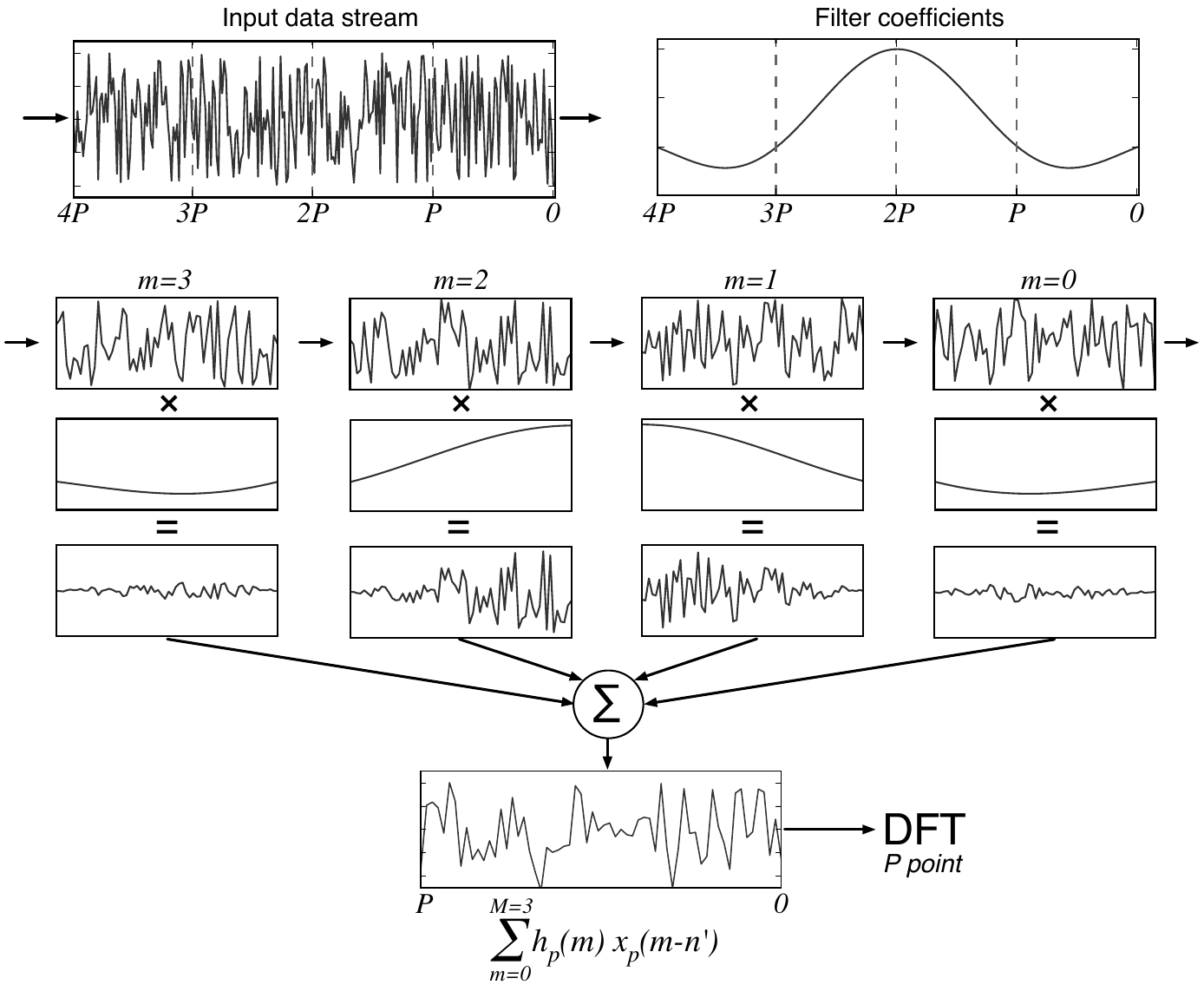}
 \caption{Graphical representation of a polyphase filterbank, with $P=64$ and $M=4$ polyphase taps. Data are read in blocks of length $P$ until $M\times P$  samples are buffered. The data and filter coefficients are then split into $M$ taps, multiplied together, and then summed over taps. After this, a $P$-point DFT is computed and another $P$ input samples are read.  \label{fig:pfb_chart}}
\end{figure}

Two different representations of PFB spectrometers are given in \fref{fig:pfb_fir_fft} and \fref{fig:pfb_chart}. \Fref{fig:pfb_fir_fft} shows a block diagram of a polyphase FIR frontend preceding an FFT. The commutator splits the input into $P$ branches, feeding a different `phase' of the input signal to each of the polyphase sub-filters. That is, the commutator applies a  $z^{-p}$ delay on each branch before a $\downarrow P$ downsampling. 

Fig.~9 shows the action of a PFB FIR frontend on a data stream in several stages. An input signal $x(n)$ of length $M\times P$ is first multiplied by filter coefficients $h(n)$. The data are then split into $M$ blocks of length $P$, and summed over $M$ taps. After this, a DFT of length $P$ is applied to form a filterbank, followed by squaring and time-averaging to compute the PSD.

A simple PFB implementation in Python is given at \url{https://github.com/telegraphic/pfb_introduction}. Provided alongside this code is an annotated interactive notebook that provides further documentation and explanation of the PFB technique. High performance codes for radio astronomy application are detailed in Refs.~\refcite{jayanth2014} and \refcite{adamek2016}.

\subsection{Zoom modes}

The output of each channel of a DFT-based filterbank is a critically-sampled quadrature time stream of its own right. %(before the signal is squared, that is). 
Higher spectral resolution can be achieved by passing the output of a `coarse' first-stage filterbank channel into a secondary DFT to apply finer channelization, after which the samples may be squared and averaged to compute the PSD. Spectrometers that employ this approach are known as \emph{zoom spectrometers}. 

An extension of the zoom spectrometer can be used to efficiently develop filterbanks of many millions of channels. The second-stage filterbank in a zoom spectrometer only needs to run at $1/N$ the rate of its first-stage filterbank. If the second-stage DFT (of length $M$) is run at the same speed as the first-stage filterbank, one can run the second-stage DFT on every first-stage channel, instead of just selecting one. The result is a filterbank with $N\times M$ channels. 

To do so requires that the output of every first-stage channel is buffered so that there are $M$ samples per channel, then data must be rearranged and fed to the second-stage DFT in channel order. This reorder can be considered a matrix transpose (also called a \emph{cornerturn}), rearranging from ($N$, $M$) to ($M$, $N$) order.

As an example, a zoom-style spectrometer with $N$=$M$=1024 has $N\times M / 2$=524,288 channels total. This approach is often used in the search for extraterrestrial intelligence (SETI)\cite{Siemion2011}, in order to achieve sub-hertz resolution over many hundreds of megahertz input bandwidth.

\section{Alternative spectrometer implementations}

\subsection{Swept spectrometer}\label{swept-spectrometer}

A \emph{swept spectrometer} uses a variable oscillator with a heterodyne
circuit and a low pass filter. The oscillator is typically varied, i.e.~swept, through a range of desired frequencies. As it is swept through the desired frequency range, the power of the low pass filter's output is measured and recorded. Most analog spectrum analyzers operate in this manner.

An advantage of swept spectrometers is that they can operate over large RF bandwidths. However, as only a fraction of the band is detected at any one time, less integration time is available per frequency channel. 
The RMS noise per channel in a swept spectrometer is $\sqrt{N}$ higher than an equivalent FTF spectrometer with $N$ channels covering the entire RF bandwidth. As such, swept spectrometers are best suited for cases where signals of interest are strong and wideband. 

\subsection{Analog filterbank}\label{analog-filter-bank}

An analog filterbank is just what its name implies: a bank (or collection) of analog filters. The analog filters are designed to pass through different ranges of frequencies. The power of each filter's output is measured and recorded, from which spectral features can be discerned. 

Analog filterbanks may offer very wide bandwidths, but design of very narrowband filters is challenging. Additionally, the input signal must be split multiple times, and each time the signal is split, its power halves. Unlike digital systems, the shape and gain of each filter may differ. For these reasons, analog filterbanks are uncommon in modern radio astronomy.

\subsection{Analog autocorrelator spectrometers}\label{sub:analog-acs}

An analog ACS uses analog circuitry to implement multipliers and propagation times through carefully constructed delay lines to implement the desired tap delays. 
An example implementation is given in Ref.~\refcite{Harris1998}.

The main advantage of the analog autocorrelator over its digital counterpart (\sref{sub:acs}) is that digitization need only take place at a rate commensurate with the averaging period of the correlator, rather than the bandwidth of the input signals. For this reason, analog ACS spectrometers are usually seen in systems that have instantaneous bandwidths of many gigahertz. Their major disadvantage is that the number of physical components required scales with the number of channels, making analog ACS systems with many channels---readily implemented by digital systems--- infeasible.

\section{Current technology}

Most modern implementations of spectrometers in radio astronomy are PFB based, and use commercially available high-speed ADCs. Over the years, ADC input bandwidth has grown from kilohertz to the gigahertz we see today. Specifications of some example high-speed ADCs that are currently available are given in \tref{tab:adcs}. 

\begin{table}[h]
	\tbl{Example high-speed ADCs that are currently commercially available. \label{tab:adcs}}
	{\begin{tabular}{c c c c c}
	\toprule
	Sample rate  & $n_{\rm{bits}}$ & Manufacturer & Part  \\
	(GS/s)  & & & \\
    \colrule
    5 &  8 & e2v & EV8AQ160 \\
	15 &  4 & Adsantec & ASNT7122  \\
	26 &  3{$^\dag$} & Analog Devices & HMCAD5831  \\
	30 &  6 & Micram & ADC30 \\
	\botrule
	\end{tabular}}
  \begin{tabnote}
  %\item [1] the first note ...
  $^\dag$ Plus overrange bit
  %\item[a] \url{http://www.adsantec.com/304-asnt7122-kma.html}
  \end{tabnote}
\end{table}
Once analog signals are digitally sampled, a variety of signal-processing platforms are available on which to implement the algorithms described earlier in this chapter:

\paragraph{Central Processing Units (CPUs),} of the type found in widely available laptop and desktop computers, are capable of processing only relatively small bandwidths, but are cheap and very easy to program. Though largely superseded in modern high-bandwidth systems, a notable example is UC Berkeley's distributed SETI@home project \citep{Anderson2002}.

\paragraph{Graphics Processing Units (GPUs)} are processors with thousands of arithmetic cores, capable of performing many trillions of operations every second.
GPU development is driven by the computer gaming industry, but over the last decade an increasing focus has been placed by GPU manufacturers on General-Purpose GPU (GPGPU) computing uses. GPUs have gained significant traction in spectrometers where large FFTs are required in order to achieve high spectral resolution\cite{Kondo2010}.

\paragraph{Field Programmable Gate Arrays (FPGAs)} are logic chips that incorporate many thousands of arithmetic cores in a fabric of programmable logic and interconnect. FPGAs excel at processing large data rates and provide low-level interfacing capabilities, allowing them to be directly connected to modern ADC chips. While an increasing number of off-the-shelf FPGA platforms are available, the needs of radio astronomers often motivate the design of custom boards. FPGAs are also relatively difficult to program, requiring specialist knowledge of their underlying hardware details  to utilize them efficiently. FPGAs have smaller quantities of memory than CPU or GPU processors, and are often used for high data-rate, coarse-resolution spectrometers \citep{Stanko2005}.

\paragraph{Application Specific Integrated Circuits (ASICs)} are custom-designed chips, with underlying circuitry dedicated to performing the operations defined by the designer. The custom nature of an ASIC makes it the most power-efficient computing platform, though this efficiency comes at the cost of large development time and effort. With the increasing performance and power efficiency of FPGAs and GPUs, ASIC development is not as prevalent in radio astronomy as it once was. However, ASICs may still be desirable for space-based spectrometers, where power efficiency is paramount\cite{hochman2014splash}.

\paragraph{Hybrid systems.} In many cases a spectrometer may be heterogeneous in nature, with different stages of processing performed on different hardware platforms.
Frequently, FPGAs are used to facilitate interfacing a high speed ADC chip with a network of CPU or GPU signal processing devices \citep{Siemion2011}. 
Spectrometers such as the \emph{VEGAS} spectrometer at the Robert C. Byrd Green Bank Telescope \citep{Prestage2015} also utilize FPGAs for ADC interfacing and coarse channelization, before signals are further filtered to a fine frequency resolution using GPUs.

\subsection{Common Infrastructure Development}

CPU and GPU processing platforms are developed by commercial entities, motivated by the non-astronomy markets. However, leveraging the latest hardware requires users to have access to flexible programming tools and software libraries. A number of open-source projects have emerged trying to serve this need.
The \textsc{GNURadio} project\footnote{\url{gnuradio.org}} provides a software environment for rapid development of CPU-based instruments for processing radio signals.
A number of radio astronomy projects have also developed generic software pipelines for streaming data between Ethernet networks, CPUs and GPUs (see, for example, \textsc{HASHPIPE}\footnote{\url{https://github.com/david-macmahon/hashpipe}},  \textsc{PSRDADA}\footnote{\url{http://psrdada.sourceforge.net}} and \textsc{Bifrost}\footnote{\url{https://github.com/ledatelescope/bifrost}}).

FPGA platforms are expensive to design and manufacture. For this reason, radio-astronomy groups such as the Collaboration for Astronomy Signal Processing and Electronics Research (CASPER\footnote{\url{https://casper.berkeley.edu}}) have developed a variety of general-purpose FPGA-based platforms that can interface with a suite of connectorized ADC cards.
CASPER provides a variety of open-source software tools and libraries with the aim of simplifying FPGA programming and enabling straightforward upgrading of instruments when newer, more capable hardware becomes available.

\section{Acknowledgements}

D. Price thanks J. Moran for thorough discussion and debate on the virtues of polyphase filterbanks; J. Bunton for clarifying the history of PFBs in radio astronomy; J. Hickish for his contribution to the technology overview; G. Hellbourg and S. Sadasivan for their input; and, D. Werthimer and A. Wolszczan for providing the opportunity and motivation to develop this chapter. Thanks also to the CASPER collaboration for sharing their extensive knowledge with the wider community. All figures have been made available by the CASPER collaboration under the CC-BY license\footnote{\url{http://creativecommons.org/licenses/by/4.0/legalcode}}.

\bibliographystyle{ws-rv-van}
\bibliography{spectrometers,references}

\end{document}